\documentclass[sigconf]{acmart}
\AtBeginDocument{%
  \providecommand\BibTeX{{%
    \normalfont B\kern-0.5em{\scshape i\kern-0.25em b}\kern-0.8em\TeX}}}

\setcopyright{acmcopyright}
\copyrightyear{2022}
\acmYear{2022}
\acmDOI{XXXXXXX.XXXXXXX}
\usepackage{multirow}

\acmConference[MM '22]{Proceedings of the 30th ACM International Conference on Multimedia}{October 10--14, 2022}{Lisbon, Portugal}
\acmBooktitle{Proceedings of the 30th ACM International Conference on Multimedia (MM '22), October 10--14, 2022, Lisbon, Portugal}
\acmPrice{15.00}
\acmISBN{978-1-4503-XXXX-X/18/06}

\usepackage{graphicx}

\usepackage{subcaption}
\usepackage{cleveref}   

\begin{document}

\title{Audio Deepfake Detection Based on a Combination of F0 Information and Real Plus Imaginary Spectrogram Features}

\author{Jun Xue}
\authornotemark[0]
\email{e21201068@stu.ahu.edu.cn}
\affiliation{%
  \institution{School of Computer Science and Technology, Anhui University}
  \city{Hefei}
  \country{China}
}

\author{Cunhang Fan}
\authornotemark[1]
\thanks{*Corresponding author}
\email{cunhang.fan@ahu.edu.cn}
\affiliation{%
	\institution{School of Computer Science and Technology, Anhui University}
	\city{Hefei}
	\country{China}
}

\author{Zhao Lv}
\authornotemark[0]
\email{kjlz@ahu.edu.cn}
\affiliation{%
	\institution{School of Computer Science and Technology, Anhui University}
	\city{Hefei}
	\country{China}
}

\author{Jianhua Tao}
\authornotemark[0]
\email{jhtao@nlpr.ia.ac.cn}
\affiliation{%
	\institution{NLPR, Institute of Automation, Chinese Academy of Sciences}
	\city{Beijing}
	\country{China}
}

\author{Jiangyan Yi}
\authornotemark[0]
\email{jiangyan.yi@nlpr.ia.ac.cn}
\affiliation{%
	\institution{NLPR, Institute of Automation, Chinese Academy of Sciences}
	\city{Beijing}
	\country{China}
}

\author{Chengshi Zheng}
\authornotemark[0]
\email{cszheng@mail.ioa.ac.cn}
\affiliation{%
	\institution{Institute of Acoustics, Chinese Academy of
		Sciences}
	\city{Beijing}
	\country{China}
}

\author{Zhengqi Wen}
\authornotemark[0]
\email{wenzhengqi@qiyuanlab.com}
\affiliation{%
	\institution{Qiyuan Laboratory}
	\city{Beijing}
	\country{China}
}

\author{Minmin Yuan}
\authornotemark[0]
\email{mm.yuan@rioh.cn}
\affiliation{%
	\institution{National Environmental Protection Engineering and Technology Center for Road Traffic Noise Control, Research Institute of Highway Ministry of Transport}
	\city{Beijing}
	\country{China}
}

\author{Shegang Shao}
\authornotemark[0]
\email{sg.shao@rioh.cn}
\affiliation{%
	\institution{National Environmental Protection Engineering and Technology Center for Road Traffic Noise Control, Research Institute of Highway Ministry of Transport}
		\city{Beijing}
		\country{China}
	}

\begin{abstract}

Recently, pioneer research works have proposed a large number of acoustic features (log power spectrogram, linear frequency cepstral coefficients, constant Q cepstral coefficients, etc.) for audio deepfake detection, obtaining good performance, and showing that different subbands have different contributions to audio deepfake detection. However, this lacks an explanation of the specific information in the subband, and these features also lose information such as phase.
Inspired by the mechanism of synthetic speech, the fundamental frequency (F0) information is used to improve the quality of synthetic speech, while the F0 of synthetic speech is still too average, which differs significantly from that of real speech. It is expected that F0 can be used as important information to discriminate between bonafide and fake speech, while this information cannot be used directly due to the irregular distribution of F0. Insteadly, the frequency band containing most of F0 is selected as the input feature. Meanwhile, to make full use of the phase and full-band information, we also propose to use real and imaginary spectrogram features as complementary input features and model the disjoint subbands separately. Finally, the results of F0, real and imaginary spectrogram features are fused.
Experimental results on the ASVspoof 2019 LA dataset show that our proposed system is very effective for the audio deepfake detection task, achieving an equivalent error rate (EER) of 0.43\%, which surpasses almost all systems.


\end{abstract}




\maketitle

\section{Introduction}

As a biometric technology, automatic speaker verification (ASV) \cite{reynolds1995speaker} has been widely used in various fields. However,  spoofing attacks can forge fake audio to deceive ASV. The spoofing attacks mainly include audio playback, text-to-speech (TTS) \cite{shchemelinin2013examining}, and voice conversion (VC) \cite{kinnunen2012vulnerability}. Recent spoofing attacks can easily fool existing ASV systems, which requires more advanced audio deepfake detection systems.

\vspace{0ex}
\begin{figure}[t]	
	\centering
	\includegraphics[width=0.98\linewidth]{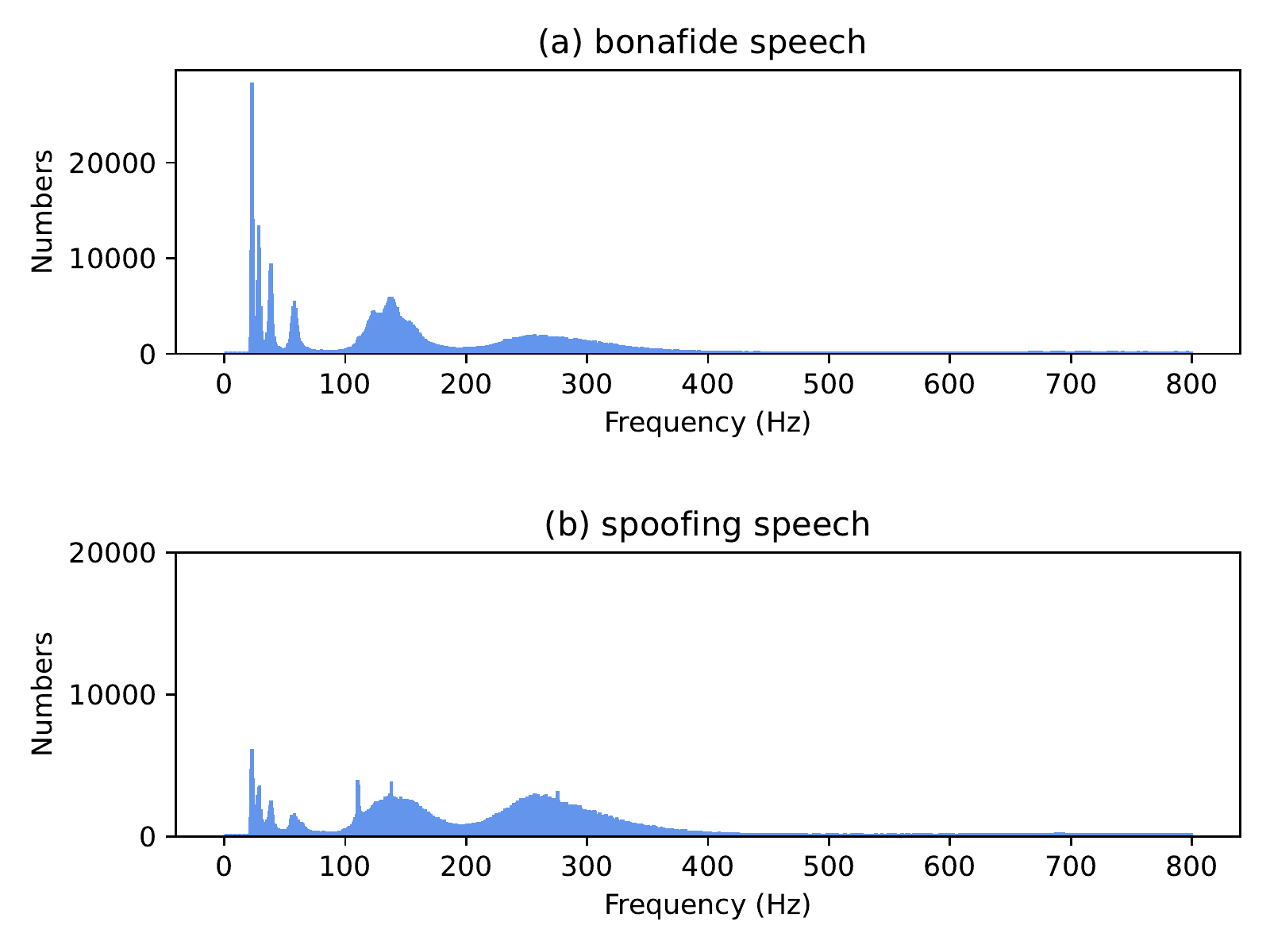}
	\caption {The F0 distribution of bonafide speech and spoofing speech in the ASVspoof 2019 LA evaluation dataset, (a) and (b) both use 7355 utterances. The horizontal axis is the frequency, and the vertical axis is the number of F0 at that frequency.}
	\label{fig:F0}
\end{figure}
\vspace{0ex}

\vspace{0ex}
\begin{figure}[t]	
	\centering
	\includegraphics[width=0.98\linewidth]{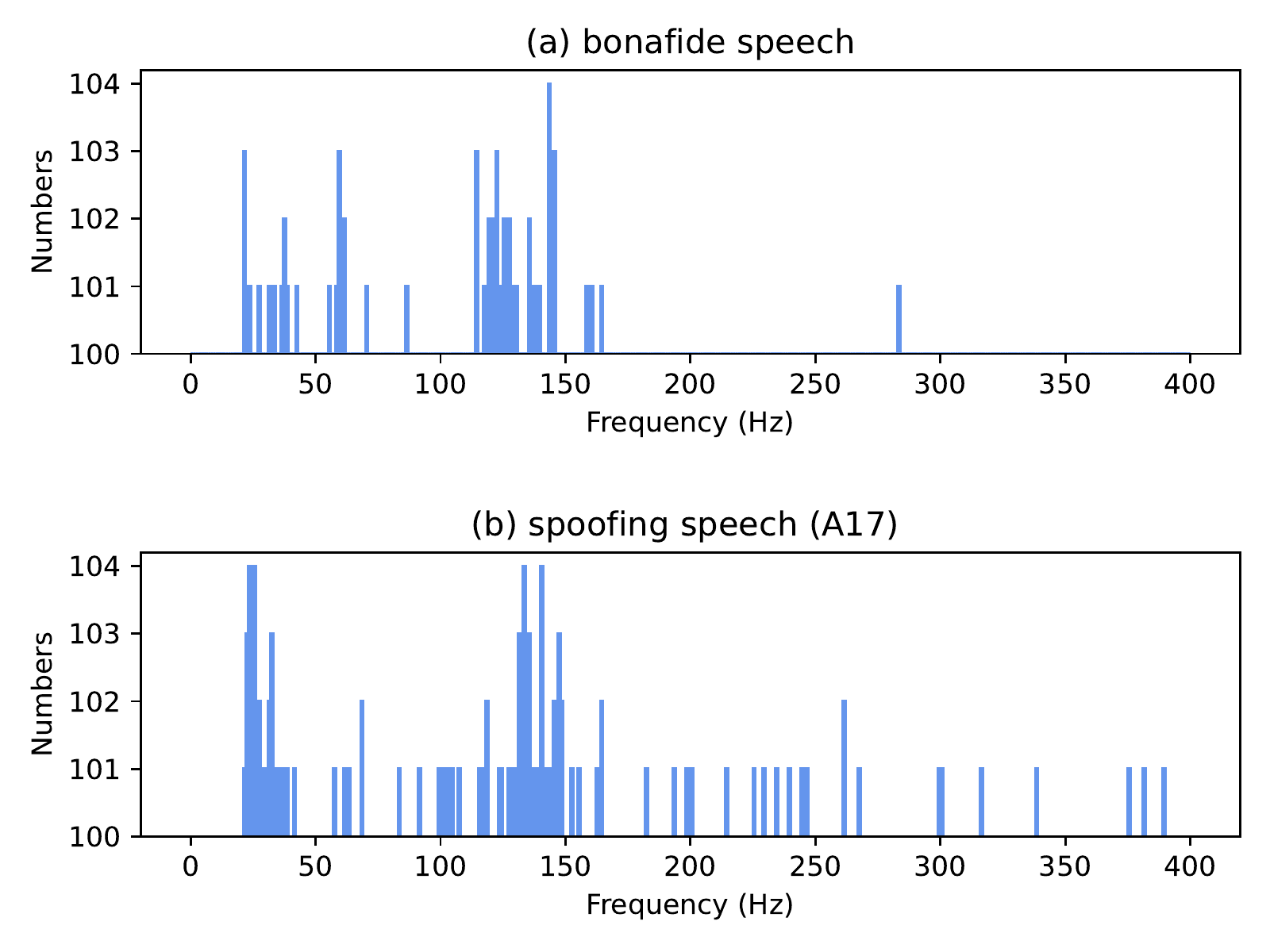}
	\caption {The F0 distribution of a single speech, where the bonafide speech is `LA$\_$E$\_$1027220.wav' and the spoofing speech is `LA$\_$E$\_$5916365.wav'.}
	\label{fig:sigle}
\end{figure}
\vspace{0ex}

To increase the security of the ASV system and reduce the risk of spoofing attacks, the ASVspoof challenge has been successfully held in 2015 \cite{wu2015asvspoof}, 2017 \cite{kinnunen2017asvspoof} , 2019 \cite{todisco19_interspeech} and 2021 \cite{yamagishi2021asvspoof}. 
And then the first Audio Deep Synthesis Detection Challenge (ADD 2022) \cite{yi2022add} was also successfully held.
To improve the performance of the audio deepfake detection system, some researchers divided the whole frequency band into different subbands and explored the impact of different frequency bands on the audio deepfake detection task. 
In \cite{chettri2020subband}, it was proposed that the whole frequency band was divided into various disjoint subbands for research. And they proposed a joint subband modeling framework that uses $n$ different subnetworks to learn specific features of subbands. 
In \cite{wang2021investigating}, a bandpass filter was used to filter or mask part of the specified frequency band, and the experimental results showed that the frequency band of 0.1-2.4KHz contains useful information to distinguish bonafide and fake speech.
In \cite{zhang21da_interspeech}, the spectrogram was divided into two parts, high-frequency and low-frequency, and good performance was achieved in the low-frequency band. The authors concluded that the low subband effectively can avoid overfitting. 
However, as for \cite{chettri2020subband, zhang21da_interspeech}, there is no basis for dividing different subbands. In addition, part of the frequency band information is lost after dividing the subbands. 

For LPS, constant Q cepstral coefficients (CQCC), and linear frequency cepstral coefficients (LFCC), these features only contain the magnitude information. Many studies have shown that the phase spectrogram also contains a lot of speech information and is important for speech quality and intelligibility \cite{paliwal2011importance, fan2020end}. 
Therefore, for the spectrogram of a speech signal, both the magnitude and phase information are indispensable.
To make full use of the phase information, researchers have made many useful attempts. For example, In \cite{xiao2015spoofing},  the group delay (GD) was applied as the phase information to acquire good performance. 
In \cite{liu2015simultaneous}, it was proposed to use the magnitude and phase information to extract super vectors, which also achieve good performance for speaker veriﬁcation anti-spoofing.
Whereas, due to the irregularity of the phase information, it cannot be used directly. It is still a challenging problem in making full use of the phase information.

In addition, for audio deepfake detection, it is particularly important to find discriminative features of bonafide and fake speech.
In the field of speech synthetic, in order to make the generated speech sound more realistic and natural, researchers have introduced many speech characteristics, such as fundamental frequency (F0), energy, and duration.
For example, in \cite{lancucki2021fastpitch}, a new text-to-speech (TTS) model was proposed. The model predicts pitch contours during inference so that the generated speech can better match the semantics of the speech. This indicates that F0 is an important speech characteristic to improve speech naturalness.
However, TTS and VC algorithms have difficulty in simulating the F0 contours of real speech. In \cite{de2012synthetic}, the authors proposed to extract features from statistical measures of pitch patterns, also achieveing good performance for audio deepfake dectetion task.
This further indicate that F0 is an important discriminative feature. Unfortunately, the F0 information, like phase information, is difficult to use directly due to the irregularity of its distribution. Therefore, how to use F0 information is also worth studing in this field.

To address the above issues, we propose an F0 subband feature for audio deepfake detection.
Fig.~\ref{fig:F0} shows the number of F0 versus frequency for real speech and fake speech in the ASVspoof 2019 LA evaluation set.
In Fig.~\ref{fig:sigle}, we choose the real and fake voices of the male voice, and show the F0 distribution at 0-400Hz.
It is well known that A17 (waveform filtering) is the most difficult type of spoofing attack to be detected, so we selected the A17 type for fake voice.
The horizontal axis is the frequency, and the vertical axis is the nunber of F0 at that frequency.
From Fig.~\ref{fig:F0} and ~\ref{fig:sigle} we can observe the following phenomena:
1) Regardless of bonafide speech or fake speech, most of the F0 is distributed between 0-400Hz;
2) In Fig.~\ref{fig:F0}, the F0 distribution of the fake speech is relatively smooth, while the F0 distribution of the real speech has obvious crests, indicating that F0 contains important distinguishing information.;
3) In Fig.~\ref{fig:sigle}, in frequency ranging from 0 to 150 Hz, the bonafide speech and spoofing speech have the similar F0 distribution, while they are quite different in frequency from 150 Hz to 400 Hz, where the fake speech has F0 in higher frequencies. Although F0 does range from 50 Hz to 450 Hz in statistics, one utterance spoken by a person cannot have a so large range of frequency in most cases.

\begin{figure}[t]	
	\centering
	\includegraphics[width=0.98\linewidth]{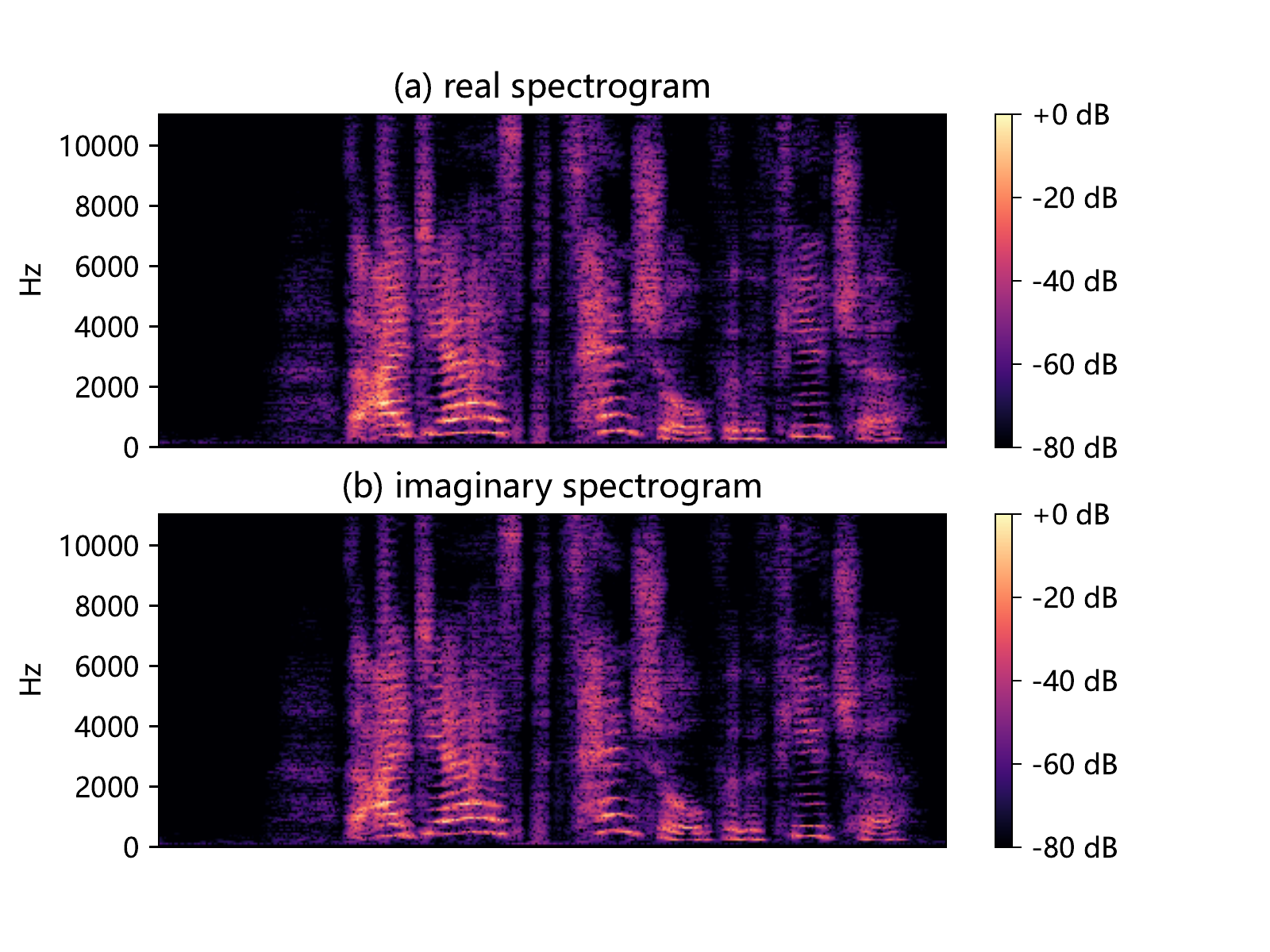}
	\caption {Real and imaginary spectrogram of `LA$\_$T$\_$1138215.wav' on the ASVspoof 2019 LA dataset.}
	\label{fig:real}
\end{figure}

Although it is relatively difficult to find the F0 distribution pattern of real speech and that of synthetic speech, as discussed above, the discriminative information of F0 can be important for the audio deepfake detection task. 
We propose to use the frequency band containing most of the F0 information as an F0 feature.
The experimental results show that the single system based on the F0 subband achieves good performance. However, the F0 subband contains only a very small part of the frequency band information, and other information of speech is also important for the detection task, such as phase spectrogram and information of other frequency bands.
Therefore, real and imaginary spectrograms are also selected as input features. Fig.~\ref{fig:real} shows the real and imaginary spectrogram of `LA$\_$T$\_$1138215.wav' on the ASVspoof 2019 LA dataset. From Fig.~\ref{fig:real}, we can see that the two magnitude spectrograms are very similar because thier only different is the phase part (Eq. \eqref{eq6} and \eqref{eq7}). Therefore, it is not necessary to fully utilize all subbands. We propose to model the low-frequency subbands of the imaginary spectrogram and the high-frequency subbands of the real spectrogram separately so that the multi-scale phase information can be fully utilized and fused to obtain the first-stage result.
Finally, the F0 subband and the first-stage results are fused to further improve the performance of the audio deepfake detection system. Experimental results on the ASVspoof 2019 LA dataset show that our proposed system is promising for the audio deepfake detection task, achieving an equivalent error rate (EER) of 0.43\%.

The rest of this paper is organized as follows. Section 2 introduces the related works. The proposed method is introduced in section 3. Experiments and results are given in section 4. Section 5 draws conclusions.

\section{Relate work}

F0 is the most basic parameter of speech, so many synthetic speech \cite{janyoi2020tonal, le2021towards} introduce F0 to improve the quality and similarity. The authors of \cite{ren2020fastspeech} have further improved the speech quality by extracting the duration, pitch, and energy in the speech waveform and using the predicted values in their inference. The literature \cite{qian2020f0} significantly improves the speech similarity by improving the autoencoder-based speech transformation to generate speech that is consistent with the target speaker F0. 
In addition, many studies \cite{patel2017significance, ge2022, yang2019significance} show that different subbands contain different discriminative information.

And about how to use the phase information has also been widely studied \cite{wang2015relative, xiao2015spoofing}. The literature \cite{chen2022fullsubnet} by taking all the magnitude spectrogram, real spectrogram and imaginary spectrogram as input to make full use of the phase information. The literature \cite{wang2017spoofing} uses modified relative phase information (MRP) and combines it with Mel frequency cepstrum coefficients (MFCC) and group delay to further improve the performance of fake speech detection system.

\section{Our proposed system}

In this section, we illustrate our proposed method for audio deepfake detection by subband fusion of F0, imaginary and real spectrograms. Recently, synthetic speech has improved speech quality by introducing F0 information, but the distribution of F0 in real speech is extremely difficult to capture so the difference between F0 is very large. Fig.~\ref{fig:F0} shows that most of the F0 is distributed at 0-400Hz, therefore, we recommend using the 0-400Hz frequency band as the F0 subband frequency. Furthermore, to introduce other frequency band information and phase information, etc. in the fake speech detection task, we propose to also model the high-frequency subband of the real spectrogram and the low-frequency subband of the imaginary spectrogram.
Finally, the performance of the audio deepfake detection system is further improved by a two-stage fusion algorithm.

\begin{figure}[t]	
	\centering
	\vspace{1ex}
	\includegraphics[width=1.0\linewidth]{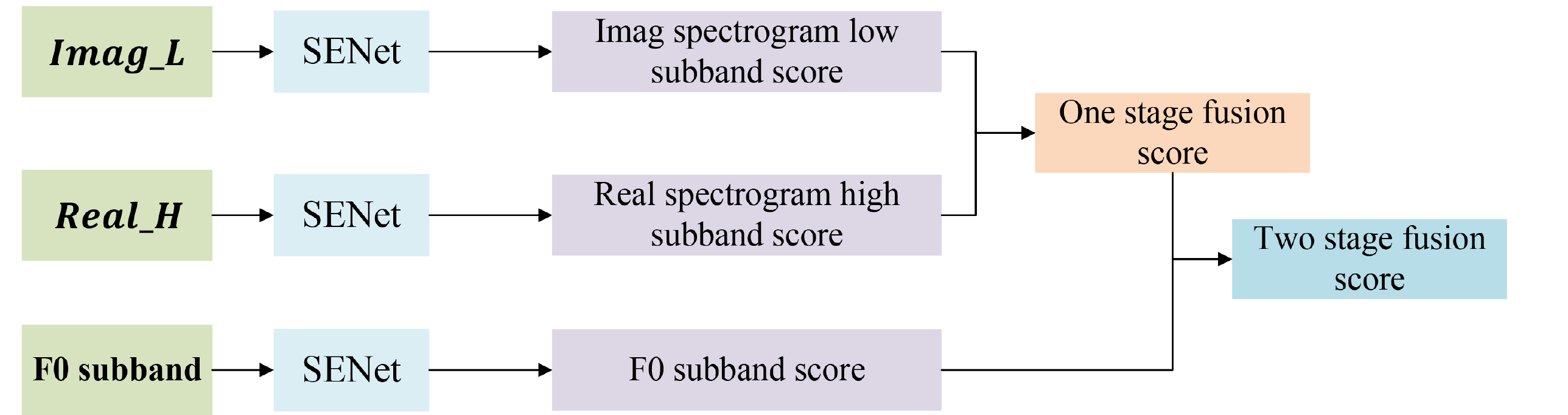}
	\vspace{1ex}
	\caption {Our proposed two-stage fusion framework.}
	
	\label{fig:Fusion}
\end{figure}

\begin{table*}[t!]
	\caption{The results of min t-DCF and EER for our proposed different frequency bands. ``F0'' represents the subbands with the widest fundamental frequency distribution. ``Rest'' represents the full bands except F0 subband. ``L'' and ``H'' denote the low and high subbands, respectively. ``Full'' represents the full frequency bands.}
	\label{tab:result1}
	\renewcommand\arraystretch{1.1}
	\setlength{\tabcolsep}{2.2mm}
	\centering
	\begin{tabular}{ccc||ccc||ccc||ccc}
		\hline \toprule
		\multicolumn{1}{c}{Systems} & t-DCF  & EER(\%)       & Systems   & t-DCF  & EER(\%) & Systems    & t-DCF  & EER(\%) & Systems    & t-DCF  & EER(\%) \\ \hline
		LPS(F0)                     & \textbf{0.0358} & \textbf{1.21} & PA(F0)    & 0.1538 & 5.26    & Imag(F0)   & 0.3387 & 12.93   & Real(F0)   & 0.3060 & 13.59   \\
		LPS(Rest)                   & 0.2450 & 11.51         & PA (Rest) & 0.2099 & 7.20    & Imag(Rest) & 0.3052 & 11.71   & Real(Rest) & 0.3145 & 11.36   \\
		LPS(L)                      & 0.0537 & 1.76          & PA (L)    & 0.1251 & 4.48    & Imag(L)    & 0.1371 & 5.44    & Real(L)    & 0.1430 & 5.84    \\
		LPS(H)                      & 0.3542 & 13.27         & PA (H)    & 0.4054 & 17.64   & Imag(H)    & 0.4596 & 18.96   & Real(H)    & 0.4598 & 26.40   \\
		LPS(Full)                   & 0.1378 & 4.80          & PA (Full) & 0.1351 & 4.67    & Imag(Full) & 0.2585 & 10.51   & Real(Full) & 0.2711 & 10.23   \\ 
		\hline \toprule
	\end{tabular}
\end{table*}

\begin{figure*}[ht!]
	\centering
	\begin{subfigure}{0.33\textwidth}
		\centering   
		\includegraphics[width=\linewidth]{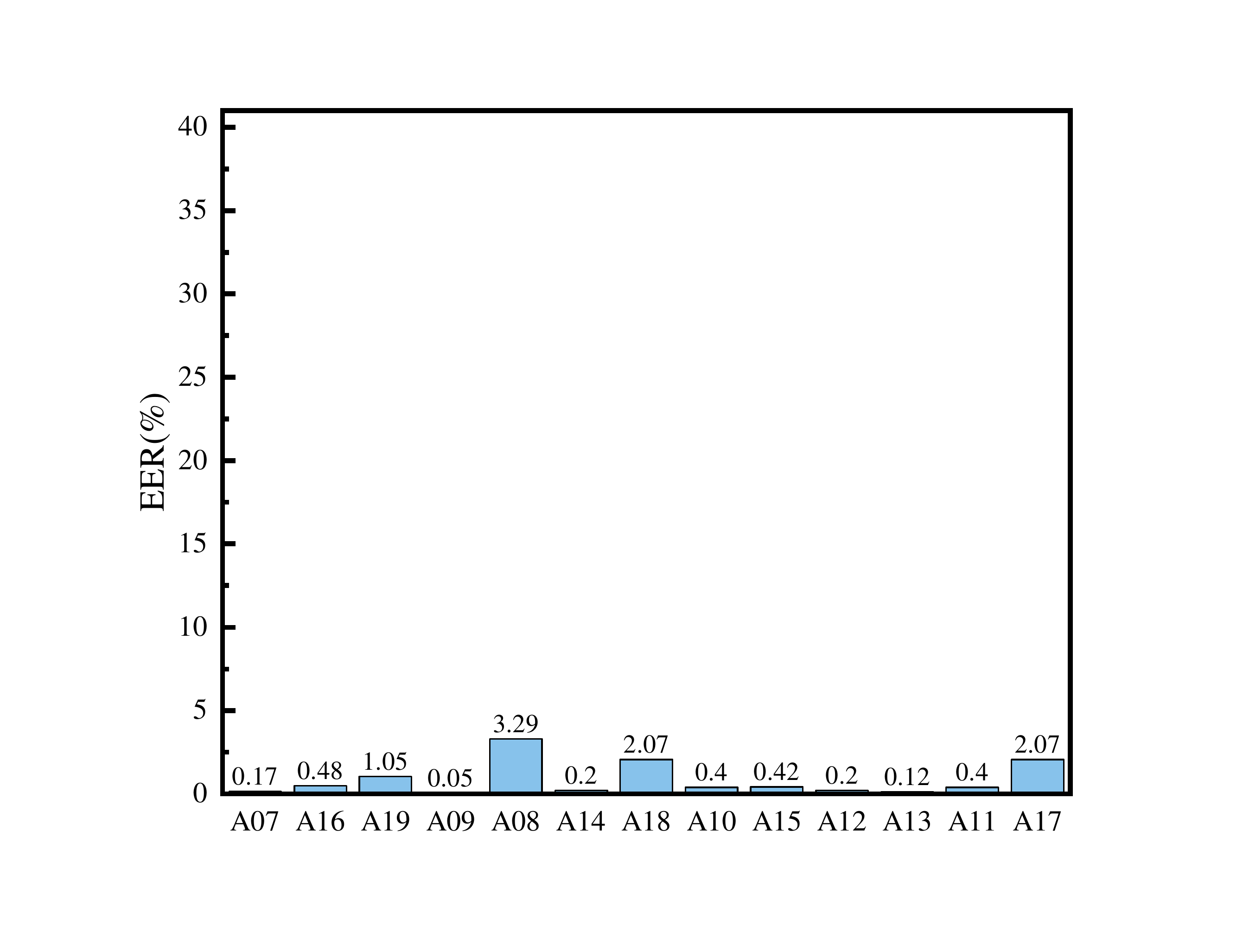}
		\caption{LPS(F0)}
		\label{fig:sub1}
	\end{subfigure}   
	\begin{subfigure}{0.33\textwidth}
		\centering   
		\includegraphics[width=\linewidth]{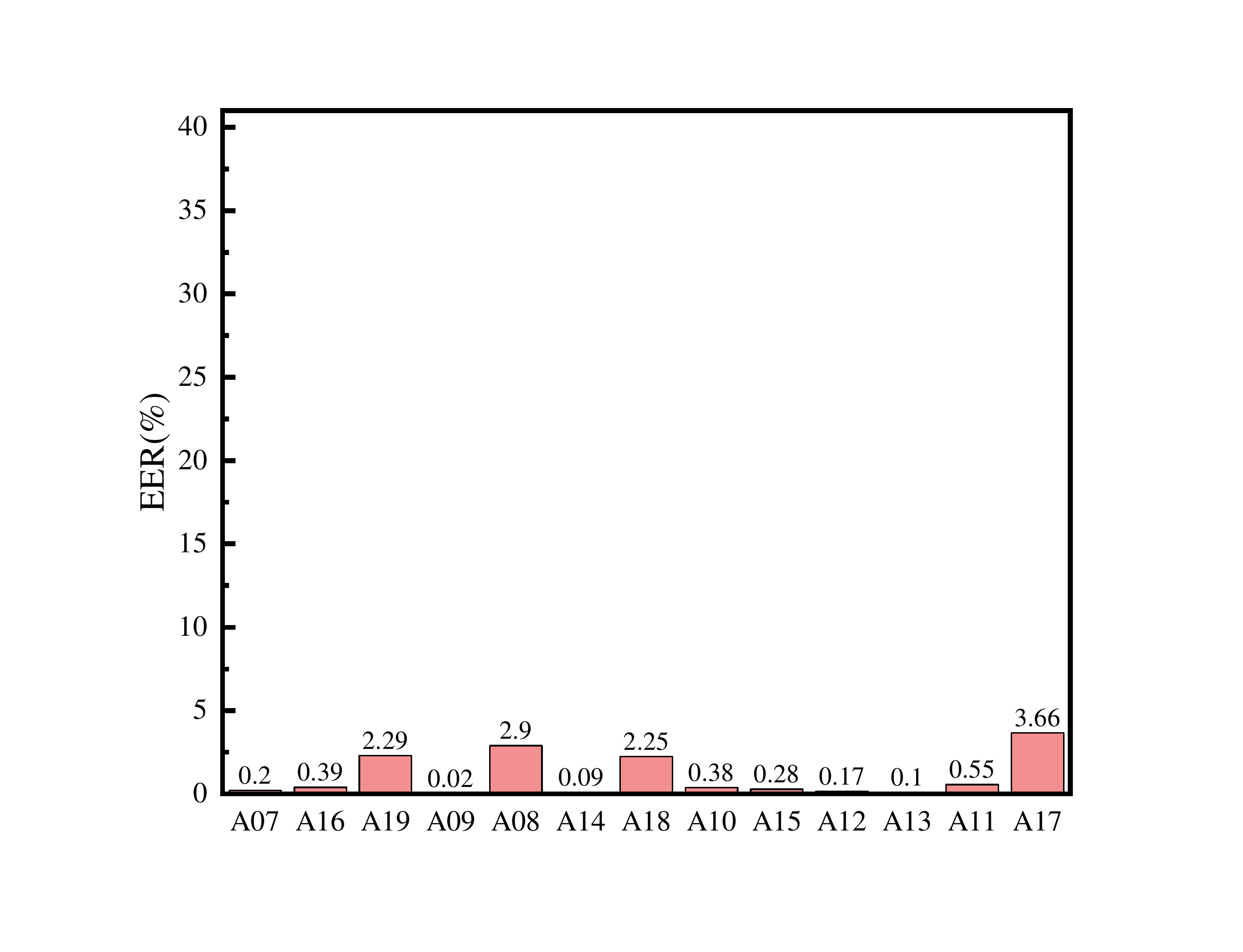}
		\caption{LPS(L)}
		\label{fig:sub2}
	\end{subfigure}
	\begin{subfigure}{0.33\textwidth}
		\centering   
		\includegraphics[width=\linewidth]{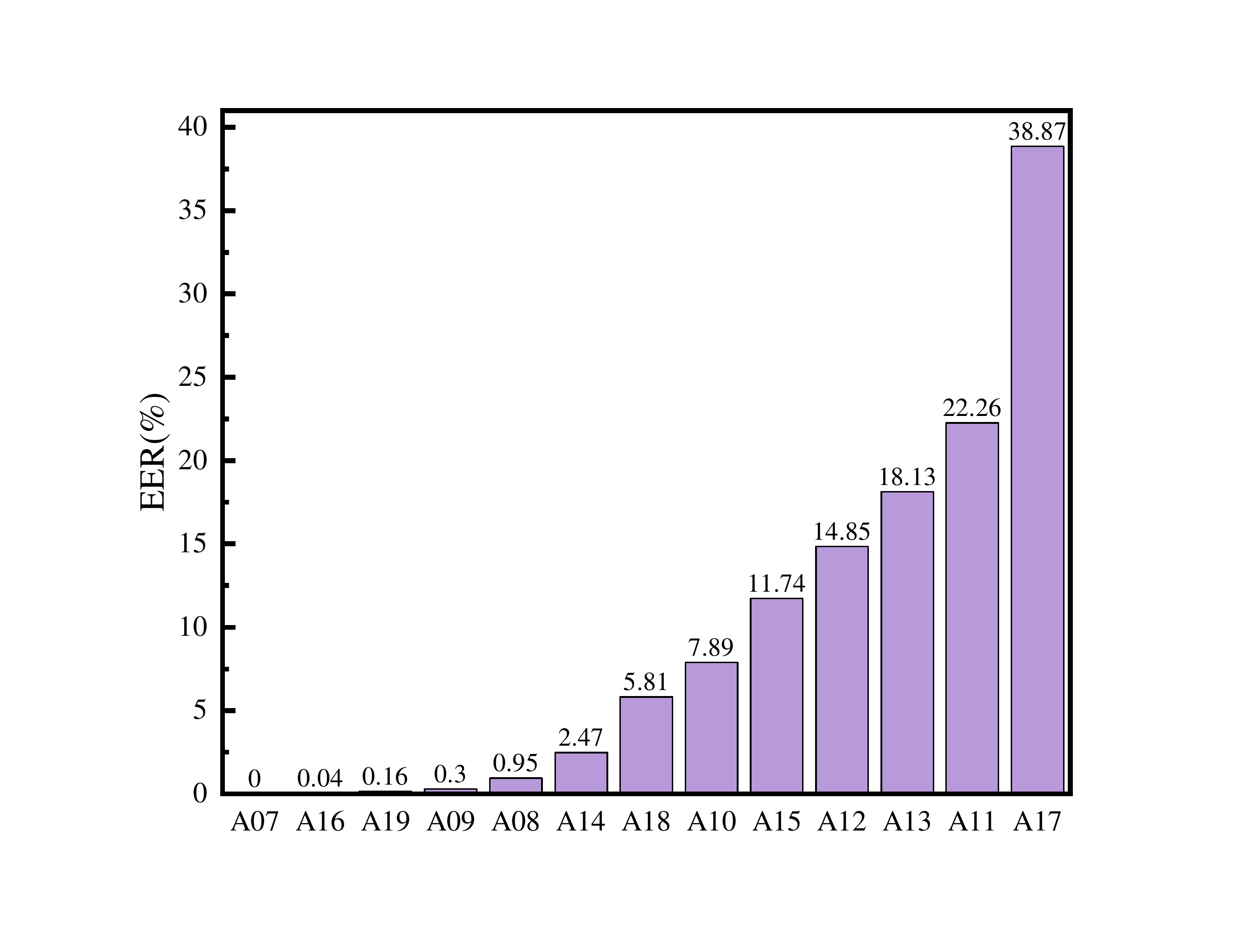}
		\caption{LPS(H)}
		\label{fig:sub3}
	\end{subfigure}	
	\begin{subfigure}{0.33\textwidth}
		\centering   
		\includegraphics[width=\linewidth]{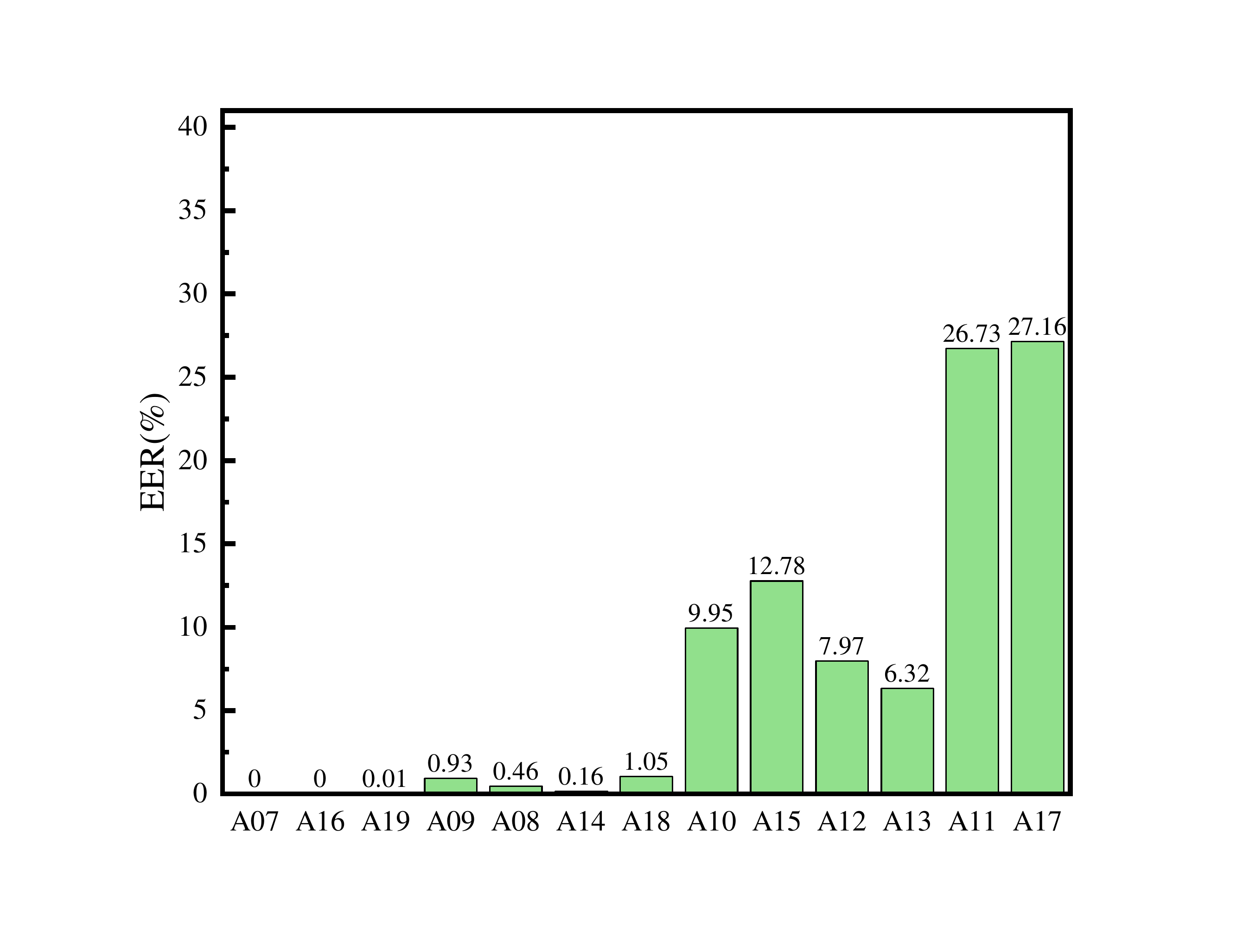}
		\caption{LPS(Rest)}
		\label{fig:sub4}
	\end{subfigure}	
	\begin{subfigure}{0.33\textwidth}
		\centering   
		\includegraphics[width=\linewidth]{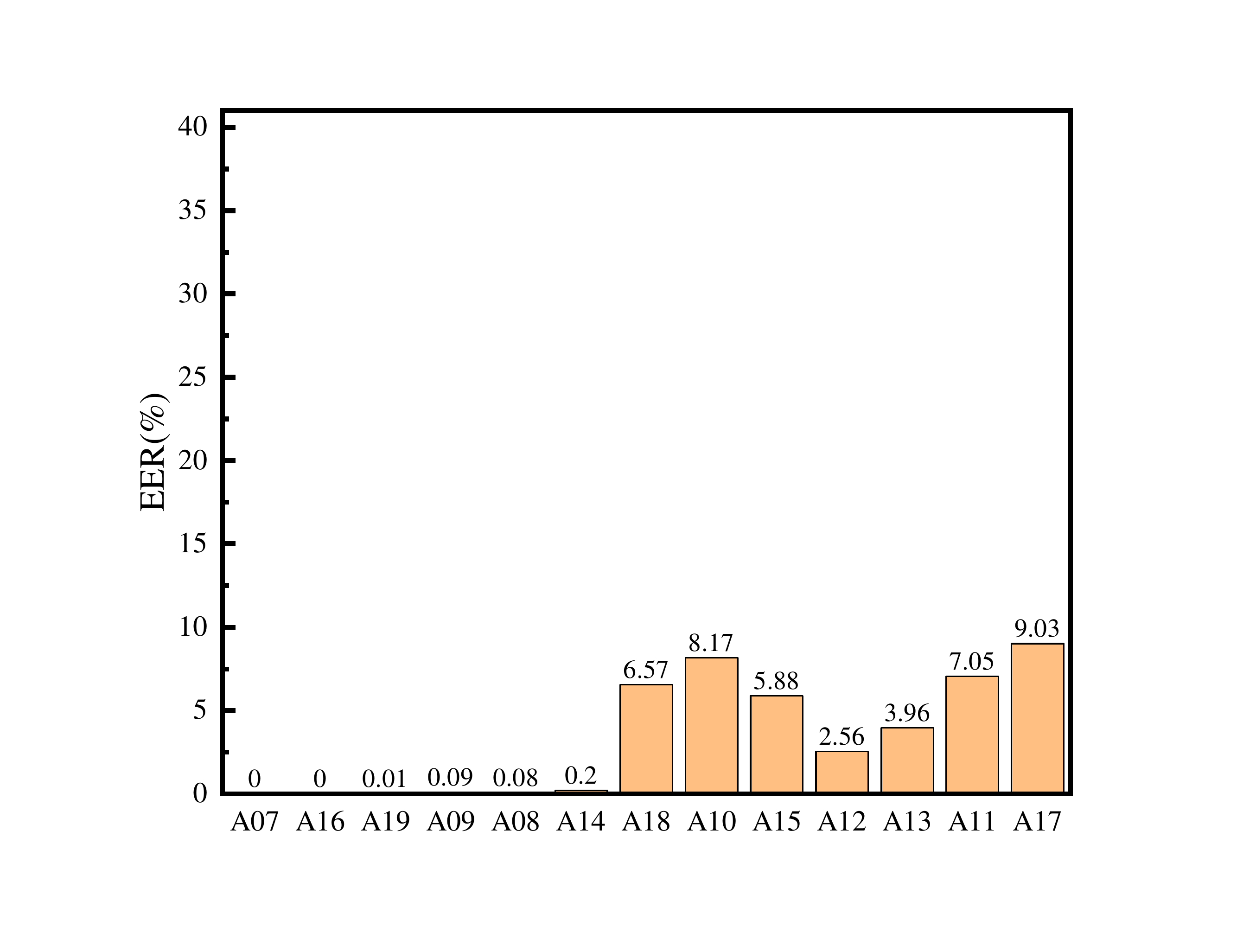}
		\caption{LPS(Full)}
		\label{fig:sub5}
	\end{subfigure}

	\caption{
		\label{fig:total}
		EER distribution of different spoofing attacks on the ASVspoof 2019 LA evaluation set. The horizontal axis is the spoofing attack type, and the vertical axis is the EER value.
	}
\end{figure*}

\subsection{F0 subband}

The proposed F0 subband is based on the LPS. Firstly, the time domain raw waveform \(\textbf{x}[k]\) is converted into time-frequency (T-F) domain by short-time Fourier transformation (STFT).
\begin{equation}
	\boldsymbol{X}_r[t,f] + i*\boldsymbol{X}_i[t,f] = STFT(\textbf{x}[k])
	\label{eq1}
\end{equation}
where \(k\) is the time index of raw waveform \(\textbf{x}[k]\). \(STFT\) means the operation of STFT. \(\boldsymbol{X}_r \in{\mathbb{R}^{{F}\times{T}}}\) and \(\boldsymbol{X}_i \in{\mathbb{R}^{{F}\times{T}}}\) are the corresponding real and imaginary part of STFT, respectively. \(t\) is the index of time frame and \(f\) is the index of frequency bin. \(F\) and \(T\) are the number of frequency bins and time frames, respectively. The full frequency bands of LPS \(\boldsymbol{LPS}_{Full}\) can be acquired as follows:
\begin{equation}
	\boldsymbol{LPS}_{Full} = log\sqrt{(\boldsymbol{X}_r)^2 + (\boldsymbol{X}_i)^2} \in{\mathbb{R}^{{F}\times{T}}}
	\label{eq2}
\end{equation}
where $\boldsymbol{LPS}_{Full}$ is the full frequency band of LPS.

From Fig.~\ref{fig:F0} we can find that most of the F0 is distributed between 0-400 Hz. Therefore, the 0-400 Hz of LPS is applied as our F0 subband \(\boldsymbol{LPS}_{F0}\).
\begin{equation}
	\boldsymbol{LPS}_{F0}=\boldsymbol{LPS}_{\text {0-400 Hz}}
	\label{eq3}
\end{equation}

\subsection{Real and imaginary spectrogram}

The F0 subband removes the phase information and most of the band information, so we consider using real and imaginary spectrograms to solve this problem. First, the  phase angle \((PA)\) and magnitude \((M)\) are expressed as follows:

\begin{equation}
	PA=\tan ^{-1}\left(X_{i} / X_{r}\right)
	\label{eq4}
\end{equation}

\begin{equation}
	M=\sqrt{\left(X_{r}\right)^{2}+\left(X_{i}\right)^{2}} \in \mathbb{R}^{F \times T}
	\label{eq5}
\end{equation}

The full frequency band of the real spectrogram \(\boldsymbol{Real}_{Full}\) and imaginary spectrogram \(\boldsymbol{Imag}_{Full}\) can be represented as follows:

\begin{equation}
	\boldsymbol { Imag }_{Full}=M * \sin (\theta)
	\label{eq6}
\end{equation}

\begin{equation}
	\boldsymbol { Real }_{Full}=M * \cos (\theta)
	\label{eq7}
\end{equation}

\((\theta)\)  is the phase angle \((PA)\) obtained above.

Inspired by \cite{zhang21da_interspeech}, we divide the real spectrogram and imaginary spectrogram into two subbands, namely low-frequency (0-4000 Hz) subband and high-frequency (4000-8000 Hz) subband. Fig.~\ref{fig:real} shows the real and imaginary spectrogram of `LA$\_$T$\_$1138215.wav' on the ASVspoof 2019 LA dataset. From Fig.~\ref{fig:real} we can see that the real and imaginary spectrogram are almost similar, so we consider that it is not necessary to fully utilize all the frequency bands of the real and imaginary spectrogram. 

Moreover, according to Eq. \eqref{eq6} and \eqref{eq7}, we can see that the phase information contained in the real spectrogram is of another scale for the imaginary spectrogram.
Therefore, we propose that using the low subband of the imaginary spectrogram and the high subband of the real spectrogram to complement each other will get better results.

\begin{equation}
	\boldsymbol{Imag}_{L}=\boldsymbol{Imag}_{\text {0-4000 Hz}}
	\label{eq8}
\end{equation}

\begin{equation}
	\boldsymbol{Real}_{H}=\boldsymbol{Real}_{\text {4000-8000 Hz}}
	\label{eq9}
\end{equation}

\begin{table}[t]
	\caption{Detailed architecture based on SENet. The channels, strides, repeat times and number of parameters are speciﬁed.}
	\label{tab:model}
	\renewcommand\arraystretch{1.1}
	\begin{tabular}{p{17mm}ccc}
		\hline \toprule Stage & Block \\
		\hline & conv2d, $16,7 \times 7$, stride $=2$, padding $=3$ \\
		Conv & BatchNorm, ReLU \\
		& maxpool, $3 \times 3$, stride $=2$, padding $=1$ \\
		\hline Layer1 & [SENet Block, 16, stride $=1$ ] $\times 3$ \\
		\hline Layer2 & [SENet Block, 32, stride $=2$ ] $\times 4$ \\
		\hline Layer3 & [SENet Block, 64, stride $=1$ ] $\times 6$ \\
		\hline Layer4 & [SENet Block, 128, stride $=2$ ] $\times 3$ \\
		\hline \multicolumn{2}{c}{ Global Average Pooling, A-Softmax } \\
		\hline \toprule
	\end{tabular}
\end{table}

\(\boldsymbol{Imag}_{L}\) is the low-frequency subband of the imaginary spectrogram, and \(\boldsymbol{Real}_{H}\) is the high-frequency subband of the real spectrogram.

\begin{table}[]
	\caption{The results of min t-DCF and EER for our proposed first stage fusion systems.  ``+'' denotes the fusion operation.}
	\label{tab:result2}
	\renewcommand\arraystretch{1.1}
	\setlength{\tabcolsep}{4.5mm}
	\centering
	\begin{tabular}{p{33mm}ccc}
		\hline \toprule
		\multicolumn{1}{c}{Systems}  & t-DCF   & EER(\%)  \\ \hline
		\multicolumn{1}{c}{LPS(F0+Rest)}                & 0.0186  & 0.59               \\
		\multicolumn{1}{c}{LPS(L+H)}                     & 0.0375  & 1.41      \\
		\multicolumn{1}{c}{PA(F0+Rest)}                  & 0.0825  & 3.03       \\
		\multicolumn{1}{c}{PA(L+H)}                      & 0.0860  & 3.16        \\
		\multicolumn{1}{c}{Imag(F0+Rest)}                & 0.1621  & 6.59     \\
		\multicolumn{1}{c}{Imag(L+H)}                    & 0.1406  & 4.25      \\
		\multicolumn{1}{c}{Real(F0+Rest)}                & 0.1736  & 7.20      \\
		\multicolumn{1}{c}{Real(L+H)}                    & 0.1676  & 5.01      \\
		\multicolumn{1}{c}{Imag(F0)+Real(Rest)}        & 0.1658  & 7.15     \\
		\multicolumn{1}{c}{Imag(L)+Real(H)}            & 0.1644  & 4.89      \\
		\multicolumn{1}{c}{Imag(Rest)+Real(F0)}        & 0.1742  & 6.56      \\
		\multicolumn{1}{c}{Imag(H)+Real(L)}            & 0.1469  & 4.43    
		\\ \hline \toprule
	\end{tabular}
\end{table}

\begin{table*}[]
	\caption{The results of min t-DCF and EER for our proposed second stage fusion systems. }
	\label{tab:result3}
	\renewcommand\arraystretch{1.1}
	\setlength{\tabcolsep}{4.5mm}
	\centering
	\begin{tabular}{ccc||ccc}
		\hline \toprule
		\multicolumn{1}{c}{Systems}  & t-DCF   & EER(\%) & Systems                              & t-DCF   & EER(\%) \\ \hline
		
		PA(F0+Rest)+ LPS(F0)           & 0.0184 & 0.63   & Imag(F0+Rest)+ LPS(F0)               & 0.0368  & 1.18 \\
		PA(F0+Rest)+ LPS(L)            & 0.0286  & 0.89  & Imag(F0+Rest)+ LPS(L)                & 0.0477  & 1.72  \\
		PA(F0+Rest)+ LPS(F0+Rest)      & 0.0267  & 0.98   & Imag(F0+Rest)+ LPS(F0+Rest)          & 0.0501  & 1.85 \\
		PA(F0+Rest)+ LPS(L+H)          & 0.0341  & 1.30    & Imag(F0+Rest)+ LPS(L+H)              & 0.0654  & 2.52   \\
		PA(L+H)+ LPS(F0)               & 0.0226  & 0.70    &Imag(L+H)+ LPS(F0)           & \textbf{ 0.0159}   & \textbf{0.47}  \\
		PA(L+H)+ LPS(L)                & 0.0326  & 1.09    &Imag(L+H)+ LPS(L)            & 0.0271  & 0.85  \\
		PA(L+H)+ LPS(F0+Rest)          & 0.0178  & 0.58   & Imag(L+H)+ LPS(F0+Rest)      & 0.0158  & 0.48  \\
		PA(L+H)+ LPS(L+H)              & 0.0287  & 0.99    &Imag(L+H)+ LPS(L+H)          & 0.0462  & 1.54  \\

 		\hline \hline
		Systems                      & t-DCF   & EER(\%) & Systems                              & t-DCF   & EER(\%) \\ \hline 
		Real(F0+Rest) + LPS(F0)      & 0.0321  & 1.03    & (Imag(H)+Real(L)) + LPS(F0)          & 0.0165  & 0.50    \\
		Real(F0+Rest) + LPS(L)       & 0.0481  & 1.64    & (Imag(H)+Real(L)) + LPS(L)           & 0.0299  & 0.89    \\
		Real(F0+Rest) + LPS(F0+Rest) & 0.0512  & 1.68    & (Imag(H)+Real(L)) + LPS(F0+Rest)     & 0.0156  & 0.49    \\
		Real(F0+Rest) + LPS(L+H)     & 0.0664  & 2.47    & (Imag(H)+Real(L)) + LPS(L+H)         & 0.0482  & 1.56    \\
		Real(L+H) + LPS(F0)          & 0.0156  & 0.50    & (Imag(L)+Real(H)) + LPS(F0)          & \textbf{0.0143}   & \textbf{0.43}    \\
		Real(L+H) + LPS(L)           & 0.0278 & 0.89    & (Imag(L)+Real(H)) + LPS(L)           & 0.0266  & 0.84    \\
		Real(L+H) + LPS(F0+Rest)     &\textbf{ 0.0133}   & \textbf{0.44}    & (Imag(L)+Real(H)) + LPS(F0+Rest)     & 0.0212  & 0.66    \\
		Real(L+H) + LPS(L+H)         & 0.0448  & 1.42    & (Imag(L)+Real(H)) + LPS(L+H)         & 0.0429  & 1.40    \\ \hline \toprule
	\end{tabular}
\end{table*}

\vspace{2ex}

\subsection{F0, imaginary and real spectrogram subbands fusion}

Inspired by \cite{zhang21da_interspeech}, the squeeze-and-excitation ResNet (SENet) \cite{hu2018squeeze} is applied as the classifier of different subbands. Those different subbands are fed into each SENet to obtain their scores. In order to make full use of other frequency bands and phase information, we propose a two-stage fusion algorithm to further improve the performance of the audio deepfake detection system. Fig.~\ref{fig:Fusion} shows our proposed two-stage fusion framework. Firstly, the scores of \(\boldsymbol{Imag}_{L}\)  and \(\boldsymbol{Real}_{H}\) are fused:
\begin{equation}
	\boldsymbol{Q}_{1} = \alpha*\boldsymbol{Q}_{Imag}^L + (1-\alpha)*\boldsymbol{Q}_{Real}^H
	\label{eq10}
\end{equation}
where the $\boldsymbol{Q}_{Imag}^{L}$ and  $\boldsymbol{Q}_{Real}^{H}$ represent the scores of \(\boldsymbol{Imag}_{L}\) and \(\boldsymbol{Real}_{H}\), respectively. The fusion score of the first stage is expressed as \(\boldsymbol{Q}_{1}\). 
Thus, \(\boldsymbol{Q}_{1}\) contains most of the speech information, including other frequency band and multi-scale phase information.
$\alpha$ is the weight of the fisrt-stage fusion.

Finally, the score of F0 is fused with \(\boldsymbol{Q}_{1}\) to improve the performance of audio deepfake detection:
\begin{equation}
	\boldsymbol{Q}_{2} = \beta*\boldsymbol{Q}_{1} + (1-\beta)*\boldsymbol{Q}_{F0}
	\label{eq11}
\end{equation}
where $\boldsymbol{Q}_{2}$ is the final fusion score, $\boldsymbol{Q}_{F0}$ is the F0 subband score \(\boldsymbol{LPS}_{F0}\). $\beta$ is the weight of the second-stage fusion.

\section{Experiments and Results}

\subsection{Dataset}

The ASVspoof 2019 database contains two subsets: LA and PA. The experiments in this paper are based on the LA subset. The LA subset contains three types of spoofing attacks: TTS, VC, and audio replay. The LA subset is divided into three parts, training, development, and evaluation. The training and development sets share the same six attacks (A01-A06), including four TTS and two VC algorithms. The evaluation set includes two known attacks (A16 and A19) and 11 unknown attacks (A07-A15, A17, and A18).

The training set is used to train the model, the development set is used to select the model with the best performance, and finally, the evaluation set is used for evaluation.

In order to quantitatively evaluate the performance of different audio deepfake detection systems, in this paper, all experiments are used EER and the minimum normalized tandem detection cost function (min t-DCF) as evaluation indicators. The EER is the working point where the false rejection rate (FRR) and false acceptance rate (FAR) are equal.

%

\subsection{Experimental setup}

As for the STFT, we set the Blackman window length and hop length of short-time Fourier transform to 1728 and 130, respectively. Therefore, the dimension of the spectrogram is 865. In order to facilitate batch processing, we fix the number of the frame at 600 by truncation or concatenating. 

F0 subband in our experiment is based on LPS feature extraction. From Fig.~\ref{fig:F0} we can find that most of the F0 is distributed between 0-400 Hz. Therefore, we apply the 0-400 Hz LPS as our F0 subband. Therefore, the first 0-45 dimension is used as the F0 subband, and finally, the feature size of the F0 subband is 45$\times$600.

For the imaginary spectrogram, we choose dimensions 0-433 as low-frequency subbands; for the real spectrogram, we choose 433-865 dimensions as high-frequency subbands. Finally, the feature sizes of the low-frequency subbands of the imaginary spectrogram and the high-frequency subbands of the real spectrogram have dimensions 433$\times$600 and 432$\times$600, respectively.
In addition, we also use the phase angle (PA) as our contrast feature, and the subband division is the same as the other features.


Inspired by \cite{zhang21da_interspeech}, we use SENet34 \cite{hu2018squeeze} as deep neural network classifier in the experiment. Table~\ref{tab:model} illustrates the architecture of SENet34, including strides, channels, and repeat times. In addition, during the experiment, we use Adam as the optimizer, and the parameters of the optimizer are set to: $\beta_{1}=0.9$, $\beta_{2}=0.98$, $\epsilon=10^{-9}$ and weight decay $10^{-4}$. The number of the epoch is 32.

$\alpha$ and $\beta$ are the weight coefficient of our two-stage fusion algorithm, both set to 0.5.

\vspace{1ex}

\begin{table}[!t]
	\caption{EER and t-DCF of single systems and primary systems based on the top performance of ASVspoof 2019 LA dataset.}	
	(a) Single systems
	\label{tab:result4a}
	\renewcommand\arraystretch{1.1}
	\centering
	\begin{tabular}{p{50mm}cc}
		\hline \toprule System & t-DCF & EER\% \\ \hline
		CQCC+GMM (B1) & $0.2366$ & $9.57$ \\
		LFCC+GMM (B2) & $0.2116$ & $8.09$ \\
		LFCC-LCNN \cite{lavrentyeva2019stc} & $0.1000$ & $5.06$ \\
		FFT-LCNN \cite{lavrentyeva2019stc} & $0.1028$ & $4.53$ \\
		LFCC-Siamese CNN \cite{Lei2020} & $0.0930$ & $3.79$ \\
		FFT-LCGRNN \cite{gomez} & $0.0776$ & $3.03$ \\	
		RW-Resnet \cite{ma21d_interspeech} & $0.0820$ & $2.98$ \\	
		Ling et al. \cite{ling21_interspeech} & $0.0510$ & $1.87$ \\		
		FFT-L-SENet \cite{zhang21da_interspeech} & $0.0368$ & $1.14$ \\
		AASIST \cite{gomez} & $0.0347$ & $1.13$ \\
		
		\hline
		LPS(F0)  \textbf{(ours)} & $\textbf{0.0358}$ & $\textbf{1.21}$ \\
		\hline \toprule
	\end{tabular}
	
	\vspace{4ex}	
	(b) Primary systems
	\label{tab:result4b}
	\renewcommand\arraystretch{1.1}
	\centering
	\begin{tabular}{p{50mm}cc}
		\hline \toprule System & $\mathrm{t}$-DCF & EER\% \\
		\hline T05 \cite{todisco19_interspeech} & $0.0069$ & $0.22$ \\
		T45 \cite{lavrentyeva2019stc} & $0.0510$ & $1.84$ \\
		T60 \cite{chettri2019ensemble} & $0.0755$ & $2.64$ \\
		GMM fusion \cite{Tak} & $0.0740$ & $2.92$ \\
		T24 \cite{todisco19_interspeech} & $0.0953$ & $3.45$ \\
		T50 \cite{Yang} & $0.1671$ & $3.56$ \\ \hline
		(Imag(L)+Real(H)) + LPS(F0)  \textbf{(ours)} & $\textbf{0.0143}$ & $\textbf{0.43}$ \\
		\hline \toprule
	\end{tabular}
\end{table}

\subsection{Experimental results}
\subsubsection{Effectiveness of F0 subband}\

Table~\ref{tab:result1} shows the minimum t-DCF and EER results of different frequency bands. ``F0'' represents the subbands with the widest fundamental frequency distribution. ``Rest'' represents the full bands except F0 subband. ``L'' and ``H'' denote the low and high subbands, respectively. ``Full'' represents the full frequency bands. 
Fig.\ref{fig:total} shows the detailed performance of LPS in different attacks of the evaluation set.

From Table~\ref{tab:result1} and Fig.\ref{fig:total}, we can observe the following aspects:

1) In the LPS feature, the EER of the F0 subband is 1.21\%, and the EER of the low subband is 1.72\%. We believe that the identification ability of LPS low subband is mainly derived from the F0 subband. This is because the low subband (0-4000Hz) has ten times frequency band information than the F0 (0-400Hz), but in the audio deepfake detection, The performance in the task is much worse. The experimental results show that the F0 subband is an important identification feature.

2) For the PA, real and imaginary spectrogram features, the performance of modeling their subbands individually is poor. This may be due to the fact that these features all contain phase information and the irregularity of the phase information leads to their poor modeling performance individually.

3) In the Fig.\ref{fig:sub1} and \ref{fig:sub2}, we can see that the F0 sub-band feature are good for the invisible attack. For the infamous attacks such as A17 (waveform filtering), the F0 subband can still capture different information with a real speech. We believe that these synthetic speech may be the lack of details of F0 information.

4) High-frequency information is important for some attack types. Even if the features of high -frequency features are not well performed in terms of overall performance, we can see that in the Fig.\ref{fig:sub3} ,Fig.\ref{fig:sub4} and \ref{fig:sub5}, compared to the lower frequency features, these features are very effective for A07 (vocoder), A16(waveform concatenation), A19 (spectral ﬁltering) and A08 (neural waveform), which is also our motivation to combine the high-frequency feature.

\vspace{2ex}

\subsubsection{Fusion results of various frequency bands}\

Table~\ref{tab:result2} and Table~\ref{tab:result3} show the minimum t-DCF and EER results for the first and second stage fusion systems, respectively. 
  ``+'' denotes the fusion operation. From the Table~\ref{tab:result2} and Table~\ref{tab:result3} we can observe the following results:

1) The non -intersecting sub -band can complement each other. Regardless of the frequency band division method, its systems can further improve performance after fusion. For example, `LPS (F0+Rest)' can get 0.59\% EER results.

2) Phase information is indispensable. In order to verify the importance of phase, we model the phase angle (PA). After the fusion of the phase angle, its performance has been greatly improved. For example, the EER results of `PA (F0+Rest)+ LPS (F0)' and PA `(L+H)+ LPS (F0)' were 0.63\% and 0.70\%, respectively.

3) The fusion of the imag spectrogram and real spectrogram can get the best results. `Imag(L+ H)+ LPS (F0)' can obtain 0.47\% EER. After replacing this high-frequency part to the real spectrogram, `(Imag (L)+ Real (H))+ LPS (F0)' can get 0.43\% EER. We think this is the income by different scale phase information.  Alternatively, `Real(L+H) + LPS(F0+Rest)' also obtains good results, but this requires the embedding of four systems and `(Imag (L)+ Real (H))+ LPS (F0)' is clearly more competitive.
%

\subsubsection{Comparison with other systems}\

%

We compare our proposed system with other systems based on the ASVspoof 2019 LA dataset. Among them, Table~\ref{tab:result4a}a shows 10 single systems that perform well, B1 and B2 being the baseline systems on the ASVspoof 2019 LA dataset. As can be seen from Table~\ref{tab:result4a}a, our proposed single system based on the F0 subband obtains 1.21\% EER and 0.0358 t-DCF, and only a small gap with the top-ranked system. 
This is because we fix the classifier, using more advanced classifiers can further improve the performance of the F0 subband.
The experimental results show that F0 is an effective discriminant feature. This not only provides a basis for dividing frequency bands but also proposes a method to fully utilize the F0 information. 

Table~\ref{tab:result4b}b shows six primary systems that perform well, our proposed fusion system based on the F0 subband of LPS and different subbands of the real and imag spectrogram achieves 0.43\% EER, outperforming the second-ranked system among all known systems (EER1.12\%). This is because for the LFCC-Capsule Fusion System \cite{luo2021capsule}, T45 \cite{lavrentyeva2019stc}, T60 \cite{chettri2019ensemble} and Ling et al. \cite{ling21_interspeech} the features are based on the magnitude spectrogram, and for the FFT-L- SENet \cite{zhang21da_interspeech} system, whose features are based on low frequency and magnitude spectrogram, which will lead to loss of information and phase information in high frequency. Although the RW-Resnet \cite{ma21d_interspeech} and RAWNet2 \cite{tak2021end} systems are based on the original waveform without losing speech information, the original waveform is affected by many factors, and it is difficult to effectively distinguish between real and fake speech.
In addition, the T05 is obtained from 7 single systems, including 2 ResNet systems, 4 MobileNet systems, and a DenseNet system. However, the system we proposed only includes 3 single systems, so we think that the system we propose is relatively simple, and we can further improve performance by improving the classifier.

\section{Conclusions}

In this paper, we propose an audio deepfake detection method based on the F0 subband, which not only fully uses the F0 information, but also provides a new and effective basis for frequency band division. 
The F0 distribution of the synthetic speech is quite different to that of the real speech, which motivates us to use F0 information as a discrimination. However, it is difficult to model F0 alone. So we use the frequency bands containing most F0 information as the F0 subband feature.
The experimental results show that the F0 subband is an effective identification feature. In addition, to make full use of voice information, we first model the low subband of imaginary spectrogram and the high subband of the real spectrogram. 
Finally, it is fused with the F0 subband by a two-stage fusion method.
The results on ASVspoof 2019 LA dataset shows that our proposed method is very effective for audio deepfake detection, and obtaining a minimum t-DCF of 0.0143 and an EER of 0.43\%. For future work, we will consider using more speech time-variant characteristics as discriminative features, and make full use of all the effective information of speech to further improve the performance of the audio deepfake detection system.
\vspace{0ex}

\section{Acknowledgements}
This work is supported by the National Key Research and Development Program of China (No. 2021ZD0201502), the National Natural Science Foundation of China (NSFC) (No.61972437), the Open Research Projects of Zhejiang Lab (NO. 2021KH0AB06), the Open Projects Program of National Laboratory of Pattern Recognition (NO. 202200014), and the National Environmental Protection Engineering and Technology Center for Road Traffic Noise Control.

\bibliographystyle{ACM-Reference-Format}
\bibliography{sample-base}

\appendix

%
%
%
%
%
%
%

\end{document}